\documentclass[preprint,showpacs,preprintnumbers,amsmath,amssymb]{revtex4}
\usepackage{graphicx}
\usepackage{dcolumn}
\usepackage{bm}
\usepackage{color}

\begin{document}
\title{Base pair fluctuations in helical models for nucleic acids}

\author{Marco Zoli}

\affiliation{School of Science and Technology \\  University of Camerino, I-62032 Camerino, Italy \\ marco.zoli@unicam.it}

\date{\today}

\begin{abstract}
A statistical method is developed to estimate the maximum amplitude of the base pair fluctuations in a three dimensional mesoscopic model for nucleic acids. The base pair thermal vibrations around the helix diameter are viewed as a Brownian motion for a particle embedded in a stable helical structure. The probability to return to the initial position is computed, as a function of time, by integrating over the particle paths consistent with the physical properties of the model potential. The zero time condition for the first-passage probability defines the constraint to select the integral cutoff for various macroscopic helical conformations, obtained by tuning the twist, the bending and the slide motion between adjacent base pairs along the molecule stack. Applying the method to a short homogeneous chain at room temperature, we obtain meaningful estimates for the maximum  fluctuations in the twist conformation with $\sim 10.5$ base pairs per helix turn, typical of double stranded DNA helices. Untwisting the double helix, the base pair fluctuations broaden and the integral cutoff grows. The cutoff is found to increase also in the presence of a sliding motion which shortens the helix contour length, a situation peculiar of dsRNA molecules.
\end{abstract}


\maketitle

\section*{I. Introduction}

Mesoscopic Hamiltonian models provide a convenient approach for studies of the structural and mechanical properties of nucleic acids as they reduce the number of degrees of freedom required to describe the nucleotide, i.e. the unit comprising the nitrogenous base and the sugar-phosphate group in the molecule backbone \cite{pablo07,doye11,jeon14,bish14,liang14}. Among the models proposed over the last decades, the Peyrard-Bishop-Dauxois (PBD) model \cite{pey93} stands out as it yields a simple and appealing description of the main forces at play in the DNA molecule in terms of a single continuous variable, accounting for the dynamics of the nucleotides on complementary strands through the stretch mode between the base pair mates. Although initially proposed to study the thermally driven DNA denaturation in the thermodynamic limit of an infinite chain, this one-dimensional model with first neighbors interactions has been later applied in different contexts to study DNA thermodynamical and flexibility properties, bubble statistics and dynamics both in short and kilo-base long sequences \cite{campa98,zdrav06,bish09,singh09,io09,io10,singh11,falo12,hando12,albu14,io14a,lak19,kalos20}. The same model has been also used to estimate the force constants of RNA chains and DNA/RNA hybrids by fitting the melting temperatures of short duplexes \cite{weber13,weber19}. 

While the original PBD model, with onsite Morse potential for hydrogen bonds, predicts base pair lifetimes of open and closed states  \cite{pey08} much shorter than those inferred from proton–deuterium exchange experiments \cite{gueron85},  successive extensions of the model \cite{pey09,erp18} have added solvent potential terms which \textit{i)} enhance the threshold for base pair dissociation over the Morse plateau and \textit{ii)} introduce a re-closing barrier accounting for the hydrogen bonds that the open bases may establish with the solvent, albeit at an higher cost as bases are hydrophobic \cite{coll95}. Although such improved model potentials partly reconcile the calculated base pair lifetimes with the experimental estimates, the fact remains that the PBD model is intrinsically 1D whereas a consistent analysis of the thermodynamics and dynamical properties of DNA  should not overlook the helical structure of the molecule with its twisting and bending degrees of freedom \cite{croq00,maiti15}. Furthermore, in a realistic picture for the DNA molecule in solution both strands recombination and large amplitude base pair fluctuations should be allowed but the pair mates should not be free to go infinitely apart. Accordingly, this requires a truncation of the base pair configuration space which is in fact always performed in computational techniques for the PBD models, with or without solvent potentials, e.g. Monte Carlo and molecular dynamics simulations and Transfer Integral methods \cite{ares05,joy05}. While the truncation removes the divergence of the thermodynamic quantities otherwise encountered for chains with a finite number of base pairs,  the cutoff is a measure of the largest separation of a fluctuating base pair identified with the width of the first bound state of the PBD Hamiltonian and generally depends on the Hamiltonian parameters \cite{zhang97}. However, even for a specific sequence,
the upper bound in the base pair fluctuations is somewhat arbitrary and markedly affects the estimated melting temperature \cite{pey95,wart85}. Consequently, large discrepancies exist as for the values used in calculations of the partition function and thermodynamical properties of DNA chains  with cutoffs ranging from $\sim 30 \, \AA$  up to about ten times the average helix diameter (which is $\sim 20 \, \AA$) \cite{zhang97,singh15}. {Also, these values hold solely for 1D molecules described by the PBD ladder Hamiltonian whereas the cutoffs may differ significantly in 3D models.}

In the finite temperature path integral method for the PBD model, the upper bound in the integration over the base pair paths can be technically determined by the normalization condition  for the free particle action \cite{io11a,io14c,klein04}. Likewise, this condition holds in the path integral for the DNA helical model developed over the last years and applied to compute cyclization, distribution lengths and stretching properties of short chains \cite{io16b,io18c,io18,io18a,io18b,io16a}.
However, for the latter purpose, one may need to take a cutoff larger than the value set by the minimal normalization condition in order to include those large amplitude fluctuations which affect the flexibility of the chain.  This points to the importance of defining a rigorous physical criterion which restricts the base pair configuration space selecting a cutoff consistently with the model potential. Here we argue that the cutoff may also vary with the specific helical conformation of the molecule which, in turn, essentially depends on the three variables, i.e. twisting, bending and sliding, describing the motion of any base pair relative to its neighbor in the stack. 

Extending to the  more realistic and complex three dimensional case an idea previously applied to the PBD ladder model \cite{io20}, we notice that the base pair thermal fluctuations are an example of Brownian motion for a particle subjected to the interactions which shape the helix. 
{Following this observation we develop a novel statistical method which sets the integration cutoff independently of the melting temperature of the specific fragment.
}
Taking a single base pair of the chain as the Brownian particle, we compute its time dependent first passage probability making use of the path integral formalism and derive a general benchmark to select the proper cutoff for the base pair fluctuations. As the argument applies in principle to any base pair in the chain, the method permits to truncate the base pairs space configuration and provides a rich relation between the amplitude of the base pair fluctuations and the helical conformation. Further, the method is general enough to be applied to any 3D helical molecule for which a Hamiltonian description is feasible.

The geometrical representation for the helix together with the Hamiltonian model are outlined in Section II while the computational method for the first passage probability is presented in Section III. The results are contained Section IV and some final remarks are made in Section V.

\section*{II. Model }

\begin{figure}
\includegraphics[height=7.0cm,width=8.0cm,angle=-90]{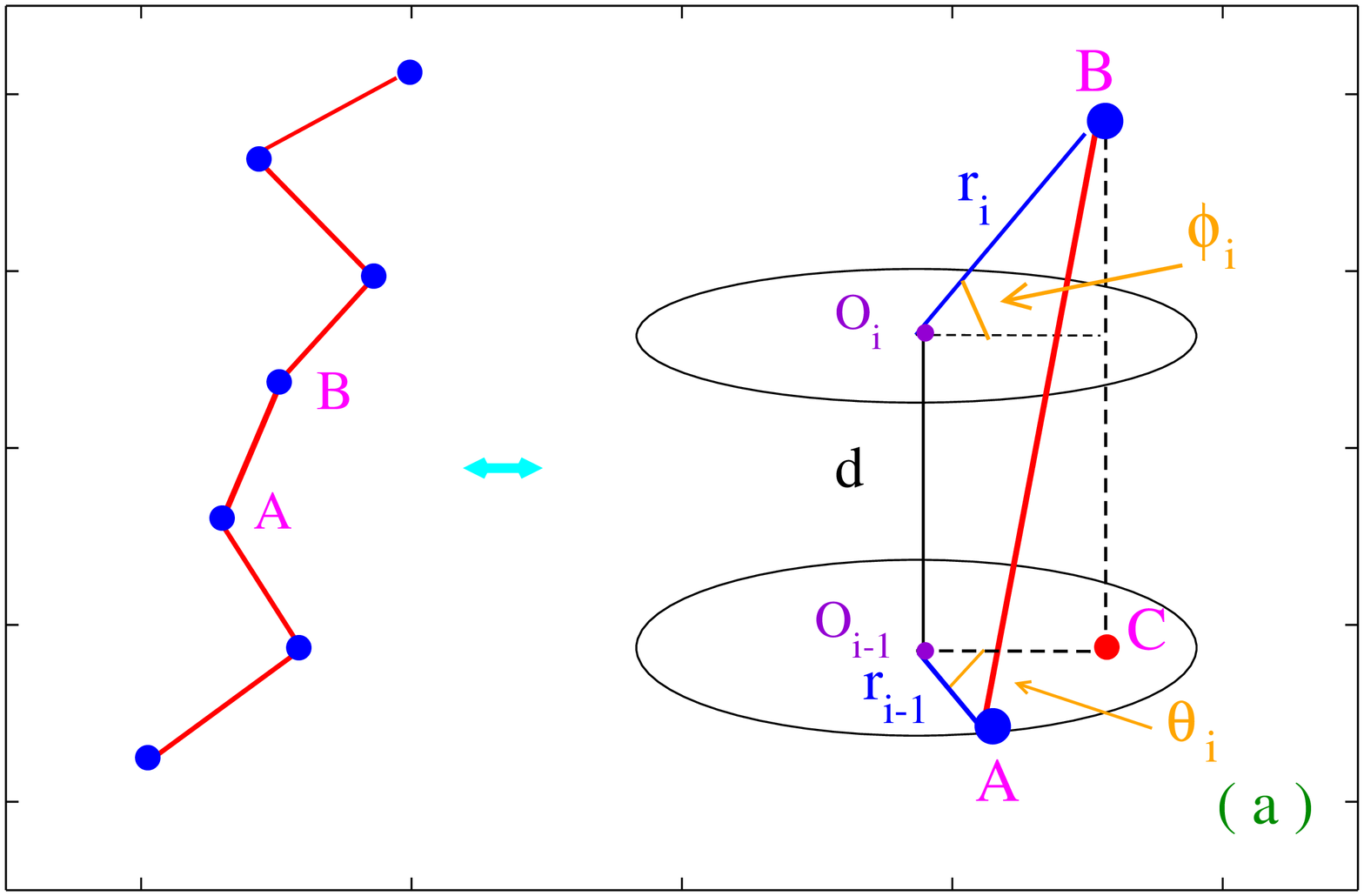}
\includegraphics[height=7.0cm,width=8.0cm,angle=-90]{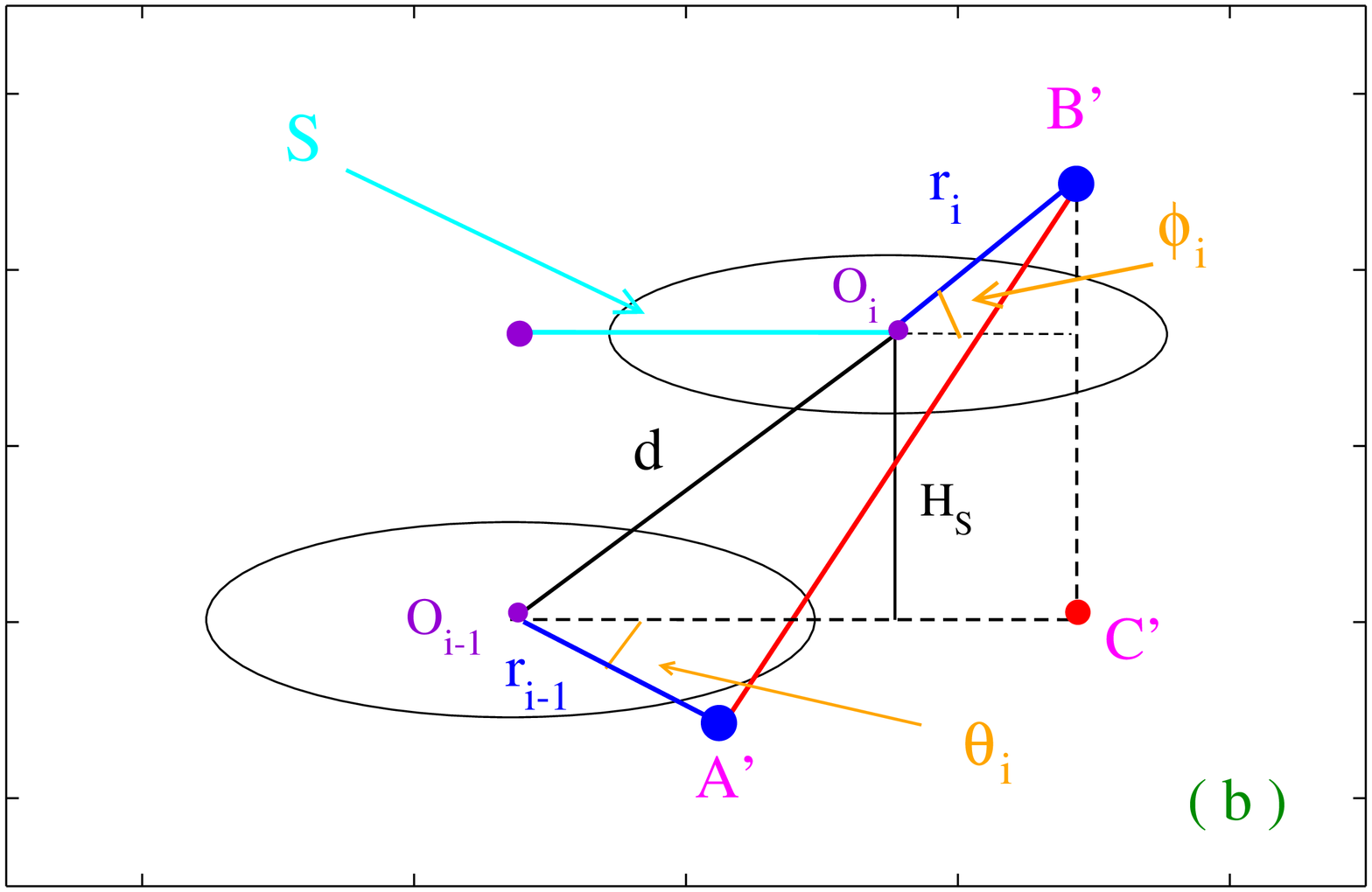}
\caption{\label{fig:1}(Color online)  
(a) Geometric model for a helical chain with $N$ point-like base pairs. $r_{i}$ is the inter-strand distance between the mates of the $i-th$ base pair. It is defined with respect to the point $O_i$ lying along the helix mid-axis. The $O_i$'s are separated by the rise distance $d$. The angles $\theta_i$ and $\phi_{i}$ define the local twist and bending between neighboring base pairs. $\overline{AB}$ is the distance between the radial displacements $r_{i}$, $r_{i-1}$. (b) Neighboring base pairs with relative sliding perpendicular to the helix mid-axis. The upper pair $(i-th)$ may go either to the left or to the right with respect to the $i-1$ pair. Note that, due to the slide $S$, the rise along the helix axis ($H_S$) is smaller than $d$.
}
\end{figure}

We consider the helical model for a linear DNA chain of $N$ point-like base pairs, with reduced mass  $\mu$, first proposed in ref.\cite{io16b} and depicted in Fig.~\ref{fig:1}(a). Essentially, the coordinate $r_{i}$ is the distance between the bases on complementary strands with respect to the mid helical axis  and it measures the base pair radial fluctuations. When $B$ overlaps $O_i$, the i-th fluctuation vanishes and the pair mates distance equals the average helix diameter $R_0$.
Neighboring base pairs along the molecule stack can be twisted and bent. In our 3D model, the angular degrees of freedom are represented by the twist $\theta_i$ and by the bending $\phi_i$ between adjacent vectors $r_{i}$ and $r_{i-1}$. In the absence of bending, the radial fluctuations occur within the ovals and the model reduces to a fixed-planes representation for the helix \cite{io11b}. Besides twisting and bending, there is a third relative motion in a dimer i.e., the sliding of the base pairs past each other, whose value depends in general on the specific pyrimidine-purine step and also on the base pair propeller twist, the latter however not accounted for by our point-like model. While the interplay of twisting, bending and sliding generates a manifold of helical conformations at the macroscopic scale \cite{calla},  the slide motion depicted in Fig.~\ref{fig:1}(b) has the main effect to shorten the rise along the helical axis and reduce the chain contour length, as found e.g., in  dsRNA  after comparison with dsDNA molecules having the same sequence \cite{gonz13,dekker14}. { In fact, A-form structures (such as dsRNA) tipically display an average absolute value of  slide larger than B-form structures although the average slide is largely dependent on the oligomer sequence.   For instance, in dsDNA, a sizeable slide is observed at pyrimidine-purine steps and also at GG/CC steps whereas \, $S$ is essentially zero at AA/TT steps \cite{calla}. Some average structural parameters found for dsRNA and dsDNA are summarized in Table~\ref{table:1}. Note that the bending angle is essentially a measure of the roll an­gle from one base pair to the next, defined in a rigid body parametrization for the local geometry of the base pair steps \cite{dick89}. Such parametrization also comprises the tilt angles which however contribute much less to the bending as they are generally smaller than the roll angles \cite{olson95}.
}

Straightforward geometrical rules permit to derive the distances between adjacent radial displacements, i.e. $\overline{AB}$ in Fig.~\ref{fig:1}(a) and $\overline{A'B'}$ in Fig.~\ref{fig:1}(b). For instance $\overline{AB}$ is given by

\begin{eqnarray}
& & \overline{d_{i,i-1}}=\, \bigl[ (d + r_i \sin \phi_i)^{2} + r_{i-1}^2 + (r_i \cos \phi_i)^2 -2 r_{i-1} \cdot r_i \cos \phi_i \cos \theta_i \bigr]^{1/2}  \, , 
\label{eq:0}
\end{eqnarray}

where $d$ is the bare rise distance in the absence of fluctuations, i.e. the bond length for beads arranged along a linear chain in the freely jointed model \cite{kovac73,busta92}. Likewise $\overline{d_{i,i-1}(S)}$ is the measure of $\overline{A'B'}$.  Note that also the twist and bending angles in Fig.~\ref{fig:1}(b) may differ from those in Fig.~\ref{fig:1}(a) due to the sliding motion.

\begin{table}
\begin{center}
 \begin{tabular}{|c| c| c|} 
 \hline
 & dsRNA & dsDNA  \\ [0.5ex] 
 \hline\hline
 $R_0$          & 24 &  20 \\ 
 \hline
 $H_S$          & 2.8 &  3.3 \\ 
 \hline
 S              & -1.48 &  0.03 \\  
 \hline
 $\bar \theta$  & $32^o$ & $34.6^o$  \\
 \hline
 $\bar \phi$  & $8.6^o$ & $3.5^o$  \\ [1ex] 
 \hline
 \end{tabular}
\end{center}
\caption{ Average structural parameters reported for A-form dsRNA and B-form dsDNA helices \cite{dekker14,olson03,wang79}. $R_0$, in units \AA, is the average helix diameter. $H_S$ and $S$, in units \AA, are respectively the rise distance per base pair and the slide as depicted in Fig.~\ref{fig:1}(b) \cite{n3}. $\bar \theta$ and  $\bar \phi$ are respectively the average twist and bending angles between adjacent base pairs.   The rise, slide, twist and bending angle may vary significantly  along the specific sequence according to the dinucleotide step \cite{oroz10,lavery14}.  }
\label{table:1}
\end{table}

From Eq.~(\ref{eq:0}), setting to zero both the intrinsic stiffness $d$ and the angular variables, one recovers the original PBD model. Later works, which have gone beyond the PBD ladder model introducing helicity without bending of the base pair planes, have assumed either finite values for $d$ and $\theta_i$  \cite{barbi99} or a non vanishing twist $\theta_i$ with zero $d$ \cite{weber06}. 

For the helical linear chain with first neighbors interactions in Fig.~\ref{fig:1}(a), the Hamiltonian reads:

\begin{eqnarray}
& &H =\, H_a[r_1] + \sum_{i=2}^{N} H_b[r_i, r_{i-1}, \phi_i, \theta_i] \, , \nonumber
\\
& &H_a[r_1] =\, \frac{\mu}{2} \dot{r}_1^2 + V_{1}[r_1] \, , \nonumber
\\
& &H_b[r_i, r_{i-1}, \phi_i, \theta_i]= \,  \frac{\mu}{2} \dot{r}_i^2 + V_{1}[r_i] + V_{2}[ r_i, r_{i-1}, \phi_i, \theta_i]  \, \, , \nonumber
\\ 
& &V_{1}[r_i]=\, D_i \bigl[\exp(-b_i (|r_i| - R_0)) - 1 \bigr]^2  \, , \nonumber
\\
& &V_{2}[ r_i, r_{i-1}, \phi_i, \theta_i]=\, K_{i, i-1} \cdot \bigl(1 + G_{i, i-1}\bigr) \cdot \overline{d_{i,i-1}}^2   \, , \nonumber
\\
& &G_{i, i-1}= \, \rho_{i, i-1}\exp\bigl[-\alpha_{i, i-1}(|r_i| + |r_{i-1}| - 2R_0)\bigr]  \, . 
\label{eq:01}
\end{eqnarray}

For the chain in Fig.~\ref{fig:1}(b),  $\overline{d_{i,i-1}}$ is replaced by $\overline{d_{i,i-1}(S)}$ in the fifth of Eqs.~(\ref{eq:01}).

$H_a[r_1]$ is taken out of the sum, as the first base pair is coupled only to the successive base pair along the chain. Being the sum of a one-particle inter-strands potential $V_{1}[r_i]$ and a two-particles intra-strand potential $V_{2}[ r_i, r_{i-1}, \phi_i, \theta_i]$, $H$ contains the essential forces which stabilize the helix \cite{proho93,olson10,metz11,maher13,onuf19}.

$V_{1}[r_i]$ is usually expressed by a Morse potential whose hard core represents the repulsive electrostatic interaction between negative phosphates on complementary strands. $D_i$ is the base pair dissociation energy which defines the Morse plateau. $b_i$ sets the potential range through its inverse value. As the bases vibrate, the pair relative distance may become even smaller than $R_0$. However, radial fluctuations have a lower bound which follows from the fact that the potential hard core reduces the statistical weight of too negative base pair contractions. Hence, by imposing the condition \, $V_{1}[r_i] \leq D_i$ \, we incorporate in the calculation only fluctuations satisfying the inequality,  $|r_i| - R_0 \geq - \ln 2/b_i$. 

A solvent potential term can be added to the Morse potential in $V_{1}[r_i]$ to enhance the height of
the energy threshold below which the base pair is closed and introduce a hump over the Morse plateau thus accounting for the re-closing barrier and possible strand recombination with the solvent. While several analytical choices have been proposed in the literature \cite{pey08}, we take in the following a widely used expression 
which well accounts for the mentioned physical effects \cite{coll95,druk01,io11b}:

\begin{eqnarray}
V_{Sol}[r_i]=\, - D_i f_s \bigl(\tanh((|r_i| - R_0)/ l_s) - 1 \bigr)
\label{eq:01a}
\end{eqnarray}

where $f_s$ is the factor which increases the energy barrier for base pair dissociation and the length $l_s$ defines the spatial range of the solvent.

The two-particle potential $V_{2}[ r_i, r_{i-1}, \phi_i, \theta_i]$ extends to the 3D helical model the peculiar stacking term chosen in the 1D PBD  model. The choice was motivated to account for the cooperativity effects in the melting transition of DNA chains, in the thermodynamic limit \cite{pey93}. 
The potential displays the stacking dependence on the angular variables and contains three parameters per dimer \, i.e., the elastic $K_{i, i-1}$ and the anharmonic force constants $\rho_{i, i-1}$, $\alpha_{i, i-1}$. Enforcing the condition \,
$\alpha_{i, i-1} < b_i$, the range of the stacking is taken larger than that of the Morse potential. Accordingly, if the inequality \, $|r_i| - R_0 \gg \alpha_{i, i-1}^{-1}$ \, holds, the hydrogen bond breaks and the local stacking interaction drops from  $ \sim K_{i, i-1}(1 + \rho_{i, i-1})$ to $ \sim  K_{i, i-1}$  thus favoring the breaking also of the adjacent base pair. This may trigger cooperatively the formation of local bubbles. 

{The Hamiltonian in Eq.~(\ref{eq:01}) differs substantially from the original PBD. First, the inclusion of the bending degree of freedom permits to address the cyclization and flexibility properties of open ends chains which are beyond the range of a ladder model. Second, the model with bending can be adapted to deal with circular molecules whereas a ladder model cannot \cite{io14b}. Third, the presence of the twist angle provides a restoring force which stabilizes the stacking potential \cite{io12}.   Precisely, if a base pair undergoes a large fluctuation $r_i$, the neighboring fluctuation $r_{i-1}$ can take only a restricted range of values so that the energy scale of $V_{2}$ remains of order $D_i$ or below. These fluctuations are those which contribute most to the partition function. Instead, in the absence of a twist, both adjacent base pair may fluctuate independently and still yield a low $V_{2}$ value.  In this sense, the stacking of the PBD model potential does not sufficiently discourage large base pair fluctuations whereas the twisted stacking $V_{2}$ in Eq.~(\ref{eq:01}) confers stability to the double stranded molecule. Importantly, it also follows that a twisted model avoids the well known divergence of the partition function found in the 1D ladder model \cite{zhang97}. Such divergence can be tackled in 1D only by truncating the real space available to base pair fluctuations. 
}

Then, Eq.~(\ref{eq:01}) models the essential interactions through five parameters which can be tweaked to account for sequence heterogeneity in chains of any length.
For short linear chains, finite size effects are relevant and should be implemented by taking open boundary conditions \cite{joy07} and parameter values at the chain ends consistent with looser hydrogen bonds and/or weaker stacking, although the strength of the latter may vary with the specific dimer \cite{zgarb14}.

Computation of the thermodynamics and mechanical properties for the model in Eq.~(\ref{eq:01}) requires integration over both the radial and the angular variables with suitable choice of the respective integration cutoffs. {In general, to perform such calculations, one has first to minimize the free energy over a broad range of possible twist angles and then select the equilibrium twist conformation, as explained in detail in ref. \cite{io17}. As a result, this procedure can be significantly time consuming even for short fragments \cite{io19}. }

However, for the task of establishing a consistent relation between model parameters and radial cutoff, the details of the base pair fluctuations over the twist and bending angles make a minor contribution. {Accordingly we chose to replace $\phi_i$ and $\theta_i$ by average values $\bar \phi$ and $\bar \theta$ which are taken as input parameters. This permits to retain the 3D nature of the model whereas the computational time is markedly reduced.}
Further, by tuning $\bar \phi$ and $\bar \theta$, one can study: \textit{i)} the relation between radial cutoff and macroscopic helical conformation, \textit{ii)} to which extent such relation is affected by the sliding motion depicted in Fig.~\ref{fig:1}(b). To this purpose we refer hereafter to the helical repeat i.e., the average number of base pair per helix turn defined by, $h=\, 2\pi / \bar \theta$  as a macroscopic measure of the twist conformation of the molecule.

\section*{III. First-passage probability for a base pair }

While DNA breathing can be tuned by binding enzymes that enhance the lifetime of the open states \cite{bonnet03,marcus13} and change the helical repeat \cite{stasiak97}, it is recognized that nucleic acids molecules display an intrinsic dynamics due to the conformational fluctuations around the average helix diameter which expose the bases to the solvent and allow regulatory proteins to access the code \cite{kalos11,wang12,marko15,lee19}.

The idea underlying the computational method is that the base pair thermal fluctuations are trajectories, providing an example of Brownian motion albeit subjected to the constraint that the base pairs are organized, at room temperature, in a stable helical structure. Accordingly, in the finite temperature path integration \cite{fehi}, the fluctuation $r_i$ in Eq.~(\ref{eq:01}) is conceived as a trajectory $r_i(\tau)$ where $\tau$  is the Euclidean time varying in the range $[0, \beta]$ and  $\beta$ is the inverse temperature. Imposing the closure condition $r_i(0)=\,r_i(\beta)$ \cite{nn1}, we can expand $r_i(\tau)$ in Fourier series around  $R_0$:

\begin{eqnarray}
r_i(\tau)=\, R_0 + \sum_{m=1}^{\infty}\Bigl[(a_m)_i \cos\bigl( \frac{2 m \pi}{\beta} \tau \bigr) + (b_m)_i \sin\bigl(\frac{2 m \pi}{\beta} \tau \bigr) \Bigr] \, ,
\label{eq:02}
\end{eqnarray}

whereby a set of coefficients defines a base pair state and measures the fluctuational amplitude.  The path integral method applied to DNA has been discussed in a number of papers, e.g. refs.\cite{io11b,io14c}, to which we refer for details.  While the expansion in Eq.~(\ref{eq:02}) holds for any base pair in the chain, the focus is now on the motion of the $j-th$ base pair identified, for instance, with the mid-chain base pair in Fig.~\ref{fig:1}(a).   For the latter, the assumed initial condition for the average pair mates separation is, $< r_j > =\,R_0$. It is remarked that, for the argument proposed hereafter, any other base pair in the chain  could have been picked besides the mid-chain one.

Due to thermal fluctuations, at any successive time $t$,  $r_j$ may exceed $R_0$ or even shrink compatibly with the physical constraint defined by the hard core Morse potential.
Accordingly, $P_j(R_0,\, t)$ is defined as the probability  that $r_j$ does not return to $R_0$ until $t$ whereas $F_j(R_0,\, t)=\, - d P_j(R_0,\, t) dt$ is the probability that the path will return for the first time to the initial $R_0$ in the time interval $dt$ past $t$. The need to introduce two time variables, $t$ and $\tau$, arises from the fact that, for a given $t$, the probabilities are given as a sum over the particle histories $r_j(\tau)$ in the time lapse  $[0, t]$ \cite{maj05}.

For the $j-th$ base pair embedded in the chain and interacting with its first neighbors via the stacking potential in Eq.~(\ref{eq:01}), we write $P_j(R_0,\, t)$ as:

\begin{eqnarray}
& &P_j(R_0,\, t)=\,   \oint Dr_{1} \exp \bigl[- A_a[r_1] \bigr] 
\cdot \prod_{i=2, \, i\neq j}^{N} \oint Dr_{i}  \exp \bigl[- A_b [r_i, r_{i-1}, \bar \phi, \bar \theta] \bigr]  \cdot \,  \nonumber 
\\
& & 
\int_{r_j(0)}^{r_j(t)} Dr_{j} \exp \bigl[- A_b[r_j, r_{j-1}, \bar \phi, \bar \theta] \bigr]  \cdot
\prod_{\tau=\,0}^{t}\Theta\bigl[r_j(\tau) - R_0\bigr] \, , \nonumber 
\\
\label{eq:03aa}
\end{eqnarray}

where the Heaviside function $\Theta[..]$ enforces the condition that $r_j(\tau)$ has to remain larger than $R_0$ for any $\tau \in [0, t]$.
This is implemented in the code by evaluating at any \, $\tau$ the amplitude of $r_j(\tau)$ in Eq.~(\ref{eq:02}) and discarding those sets of coefficients which don't comply with such condition.

The action functionals in Eq.~(\ref{eq:03aa}) are obtained by the following $d\tau$ integrals:

\begin{eqnarray}
& &A_a[r_1]=\,\int_{0}^{\beta} d\tau H_a[r_1(\tau)] \, , \nonumber 
\\
& &A_b [r_i, r_{i-1}, \bar \phi, \bar \theta]=\, \int_{0}^{\beta} d\tau H_b[r_i(\tau), r_{i-1}(\tau), \bar \phi, \bar \theta] \, , \nonumber 
\\
& &A_b [r_j, r_{j-1}, \bar \phi, \bar \theta]=\, \int_{0}^{t} d\tau H_b[r_j(\tau), r_{j-1}(\tau), \bar \phi, \bar \theta] \, ,
\label{eq:03}
\end{eqnarray}

while the measures of integrations over closed ($\oint {D}r_{i}$) and open  ($\int Dr_{j}$) trajectories are coupled via the two particle potential which connects the $i-th$ and $j-th$ base pairs along the stack. For the two measures however, different criteria should be applied to set the integration cutoffs on the radial fluctuations.

Consistently with Eq.~(\ref{eq:02}), $\oint {D}r_i$ is explicitly defined by

\begin{eqnarray}
& &\oint {D}r_{i} \equiv  \prod_{m=1}^{\infty}\Bigl( \frac{m \pi}{\lambda_{cl}} \Bigr)^2  \times \int_{-\Lambda_{i}(T)}^{\Lambda_{i}(T)} d(a_m)_i \int_{-\Lambda_{i}(T)}^{\Lambda_{i}(T)} d(b_m)_i \, , \, 
\label{eq:04}
\end{eqnarray}

where  $\lambda_{cl}$ is the classical thermal wavelength and $\Lambda_{i}(T)$ is the temperature dependent cutoff which sets the maximum amplitude for the base pair Fourier coefficients. In turn, the cutoff can be determined by using the property that the path integral measure
$\oint {D}r_{i}$ normalizes the kinetic action, i.e., 

\begin{eqnarray}
\oint {D}r_i \exp\Bigl[- \int_0^\beta d\tau {\mu \over 2}\dot{r}_i(\tau)^2  \Bigr] = \,1 \, .
\label{eq:05} \,
\end{eqnarray}

In fact, using Eqs.~(\ref{eq:02}),~(\ref{eq:04}), the l.h.s. of Eq.~(\ref{eq:05}) yields  a product over the Fourier components of decoupled Gau{\ss}ian integrals. Setting $\Lambda_{i}(T)\equiv \,{{U_i \lambda_{cl}} / {m \pi^{3/2}}}$, it is numerically found that  $U_i=\,2$ suffices to satisfy Eq.~(\ref{eq:05}). 
While this provides a formal criterion to establish a minimum $U_i$ for the closed trajectories in the chain, it is understood that larger cutoffs may be also assumed for practical tasks e.g., for the computation of flexibility properties which essentially depend on large amplitude base pair fluctuations \cite{io11a}.

As for the $j-th$ base pair, while the path expansion in Eq.~(\ref{eq:02}) can be performed, one cannot apply the normalization condition in Eq.~(\ref{eq:05}). This follows from the observation that, for any $\tau$,   $r_j(\tau)$ is defined up to $r_j(t)$ which is in fact an open trajectory for any $t < \beta$. Hence, a new criterion should be developed to estimate the integral cutoff on the amplitude of the $j-th$ radial fluctuation.

To this purpose we examine Eq.~(\ref{eq:03aa}) and ask the question: what is the probability that, at the initial time, the $j-th$ fluctuation is larger than $R_0$ ?  

From Eq.~(\ref{eq:02}), at $t=\,0$, the $j-th$ trajectory is \, $r_j(0)=\, R_0 + \sum_{m=1}^{\infty}(a_m)_j$ with the Fourier coefficients being integrated on an even domain. Accordingly, the initial probability $P_j(R_0,\, 0)$ is expected to be $ \sim 1/2$ \cite{n2}. This is the benchmark to be met in the computation of the first-passage probability as a function of time. In order to implement this criterion, we first 
truncate the Fourier integration $\int_{r_j(0)}^{r_j(t)} Dr_{j}$  in Eq.~(\ref{eq:03aa}) by a cutoff \, $\Lambda_{j}(T)=\,{{U_j \lambda_{cl}} / {m \pi^{3/2}}}$, with tunable $U_j$ and then determine the value \, $U_j$ such that $P_j(R_0,\, 0) \sim 1/2$.  In this way one selects the cutoff on the base of a physical constraint for a specific set of model parameters and for a given helical conformations.

\begin{figure}
\includegraphics[height=7.0cm,width=8.0cm,angle=-90]{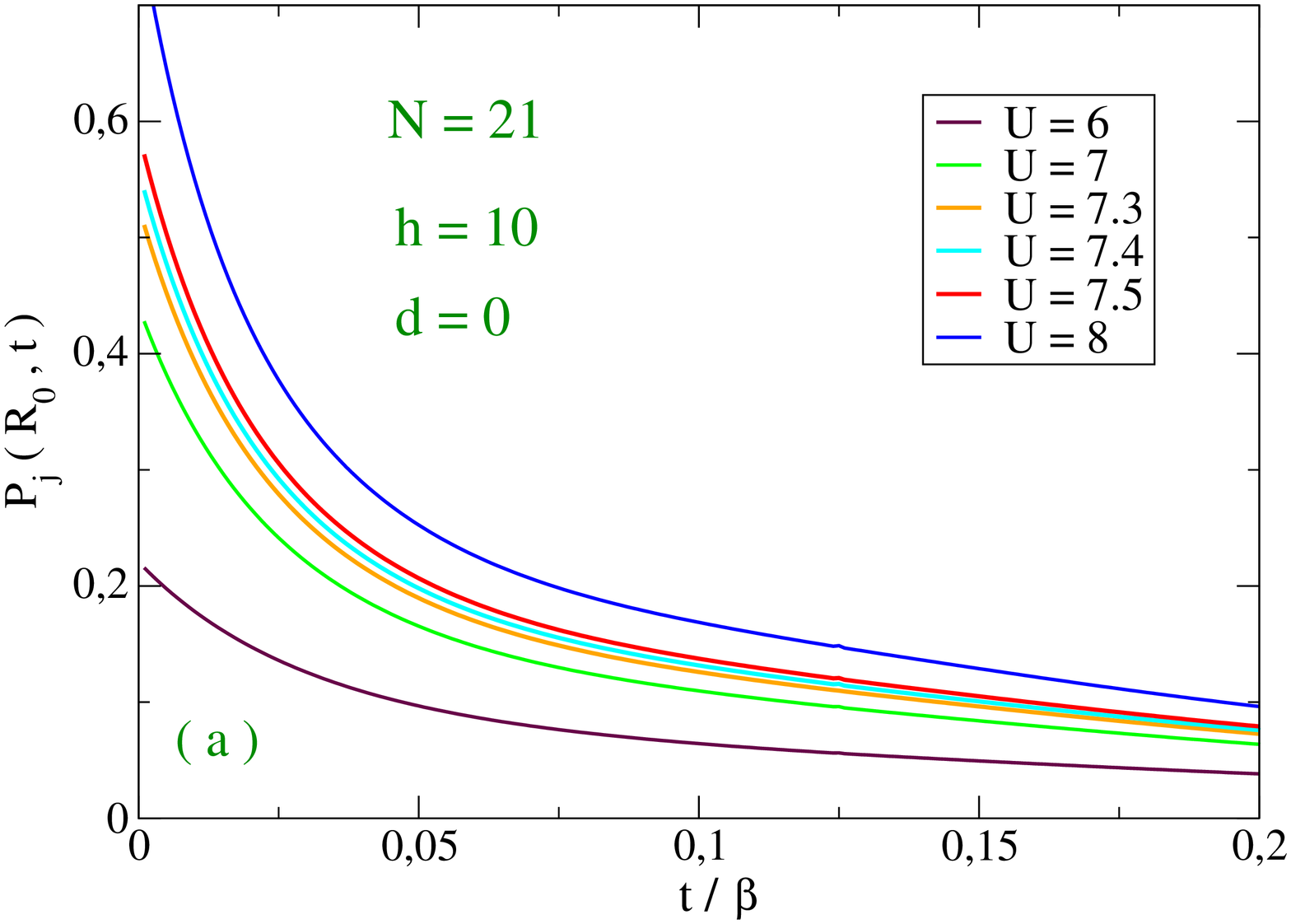}
\includegraphics[height=7.0cm,width=8.0cm,angle=-90]{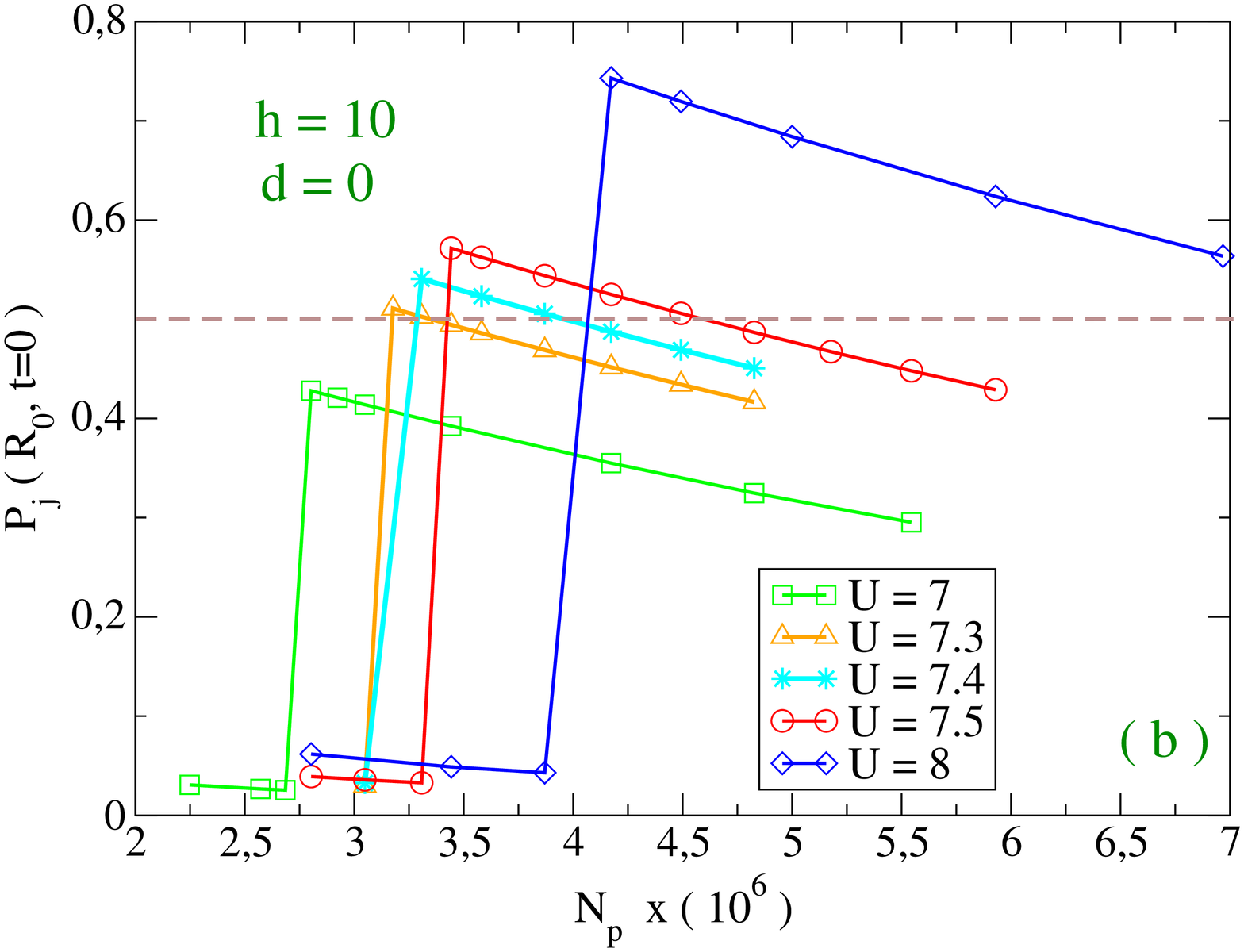}
\caption{\label{fig:2}(Color online)  
(a) First-passage probability versus time for the mid-chain base pair in equilibrium with the $N-1$ base pairs at room temperature. A homogeneous short chain with average twist and bending angles is considered. Various values for the integral cutoff ($U_j\equiv {U}$) over the radial base pair fluctuations are assumed. (b) For any integral cutoff, the zero time probability is computed as a function of the number of trajectories for the chain dimer containing the $j-th$ base pair.}
\end{figure}

\section*{IV. Results }

Our theory is first tested by setting in Eq.~(\ref{eq:03aa}) the average bending at $\bar \phi=\,6^{o}$ consistently with the indications of Fluorescence Resonance Energy Transfer studies probing the DNA bending elasticity  at short length scales \cite{kim14}. 
The average twist angle is initially taken as $\bar \theta=\,36^{o}$, which means an average helical repeat \, $h=\,10$,  close to the usual experimental value for kilo-base long DNA chains at room temperature \cite{wang79}. Although the model potential contains only first neighbors radial interactions, the base pairs are correlated along the stack due to the helical conformation. In the following calculations, a short homogeneous chain of $N=\,21$ base pairs is considered which allows for about two turns of the helix whose diameter is $R_0=\, 20 \AA$.

The potential parameters are those taken in ref.\cite{io09} namely,  $D_{i}=\,30 \, meV$, $b_{i}=\,4.2 \, \AA^{-1}$, $K_{i}\equiv K_{i,i-1}=\,60 \, meV \cdot \AA^{-2}$, 
$\rho_{i}\equiv \rho_{i,i-1} =\,1$,  $\alpha_{i}\equiv \alpha_{i,i-1} =\,0.35 \, \AA^{-1}$. This set is in line with the parameters  used by other groups  in investigations of the PBD model \cite{zhang97,campa98} and derived by fitting the experimental melting temperatures, though some discrepancies persist mostly as for the stacking force constants \cite{eijck11,io20}.  For the dimers containing the terminal base pairs, subjected to end fraying effects \cite{weber15}, the stacking parameters \, $K_{2,1}$ and $K_{N,N-1}$ are taken one half of the value assumed for the internal dimers. While this choice is arbitrary, it permits to weigh the impact of chain end effects on the base pair cutoff.  

We proceed by making, at first, the simplifying assumption to take a vanishing $d$ (like in the PBD ladder model) whose consequence will be pointed out below.

Given the set of parameters, the output of the calculation still essentially depends on \textit{a)} the cutoff $U_j$ and \textit{b)} the number of base pair trajectories included in the integrals of Eq.~(\ref{eq:03aa}). 
This is shown in Fig.~\ref{fig:2}(a) which plots the probabilities as a function of time for different $U_j$ values and in Fig.~\ref{fig:2}(b) which plots the \textit{zero time} probabilities versus the number of possible paths ($N_p$) for the dimer associated to the $j-th$ base pair.  As the time axis partitioning involves 1000 points, the \textit{zero time} corresponds to $t / \beta =\,10^{-3}$.

Fig.~\ref{fig:2}(b) highlights the dependence of $P_j(R_0,\, 0)$ on the path ensemble size and also provides the rule to select the appropriate $N_p$ for a given $U_j$. In fact, the $P_j(R_0,\, t)$'s in Fig.~\ref{fig:2}(a) are computed by assuming, for each $U_j$, the $N_p$ values which maximize the corresponding $P_j(R_0,\, 0)$'s thus ensuring that the calculation is sampling the largest set of base pair fluctuations, i.e. of molecule configurations, consistent with the model parameters.

It is found that, for \, $U_j \sim \,7.3$, the condition \, $P_j(R_0,\, 0) \sim 1/2$ is fulfilled provided that the ensemble size for the dimer fluctuations is \, $N_p \sim 3.2 \cdot 10^{6}$. 
Moreover, $N_p$ is constant for all dimers in the stack but the terminal ones as all base pairs are subjected to the same physical constraints imposed by the model potential.
Since the chain is homogeneous, the selected value is a good cutoff for all internal base pairs, thereby identified as $U_j \equiv \bar{U}$. 
Instead, for the terminal base pairs, the first passage probability remains smaller than $1/2$, e.g. $P_{j=N}(R_0,\, 0) \sim 0.018$, signalling that the average base pair separation is larger than $R_0$. This effect is ascribable to the softer stacking force constant  which causes looser bonds and fraying at the chain ends \cite{berg06}. 

\subsection*{A. Twisting}

\begin{figure}
\includegraphics[height=7.0cm,width=8.0cm,angle=-90]{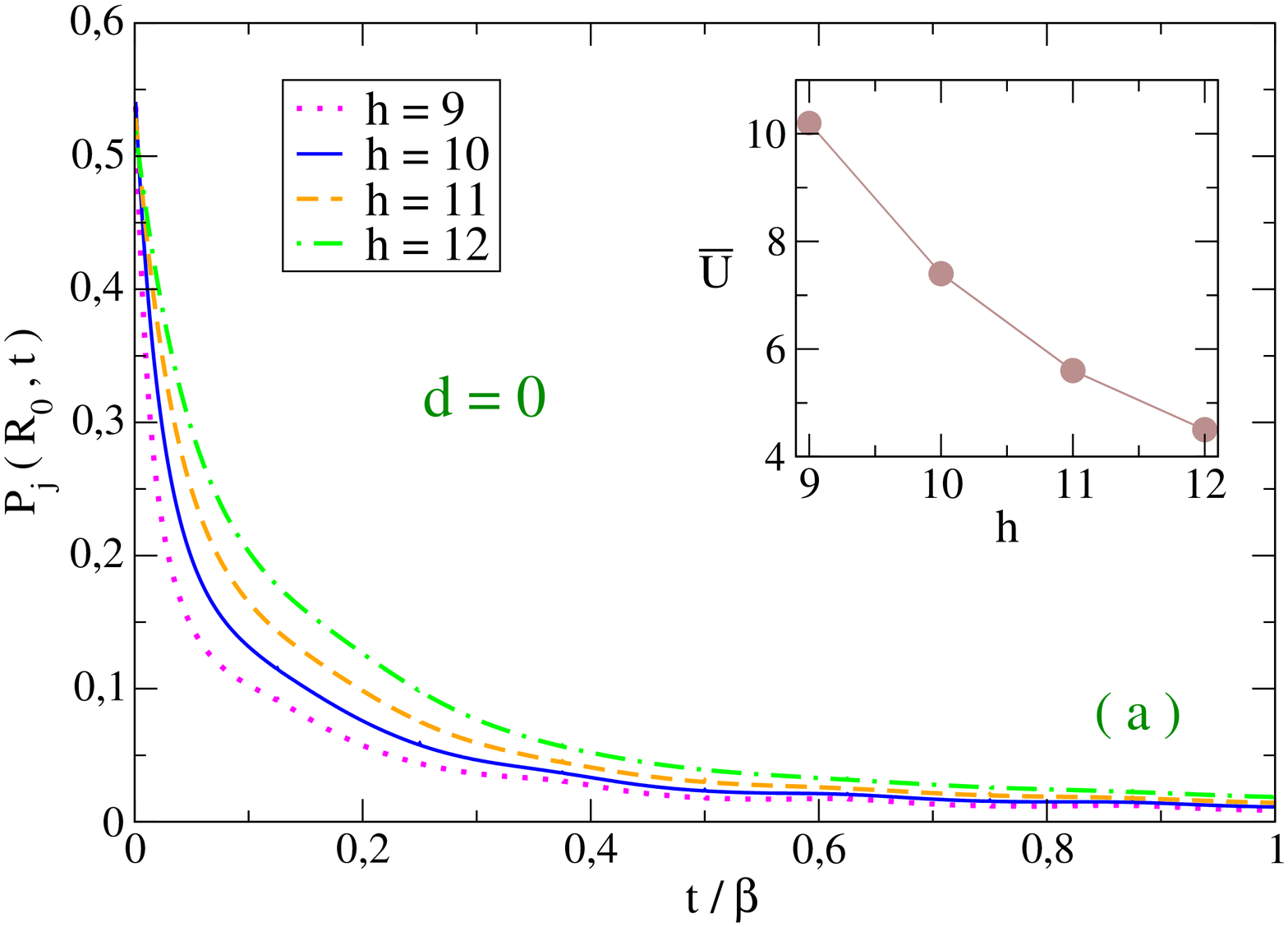}
\includegraphics[height=7.0cm,width=8.0cm,angle=-90]{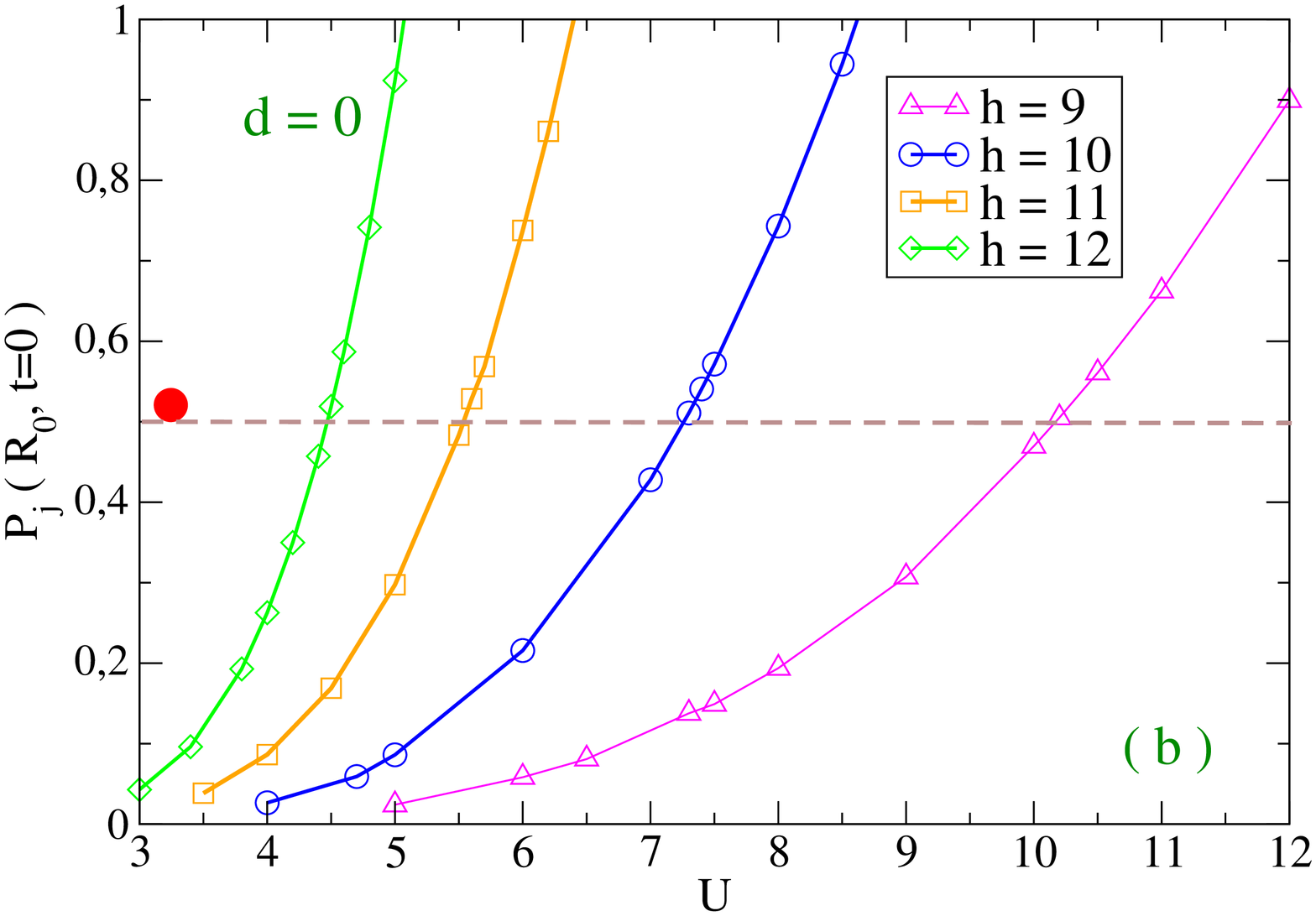}
\caption{\label{fig:3}(Color online)  
(a) First-passage probability versus time for the mid-chain base pair in equilibrium with the $N-1$ base pairs at room temperature.  The chain is taken as in Fig.~\ref{fig:2} except the average twist for which four $h$ values are considered. {The respective average twist angles $\bar \theta$  are (from top to bottom in the legend) \, $40^o, \,36^o, \, 32.7^o, \, 30^o $.}
 For each twist conformation, the probability is computed assuming the respective cutoff \={U}
(in the inset) that fulfills the initial time condition. (b) Zero time probabilities versus integral cutoff for three twist conformations. The $U$ values for which the plots intersect the dashed line, are the $\bar U$'s reported in the inset in (a).  {\color{red} $\large \bullet$ } \,  marks the cutoff found for the PBD ladder model. }
\end{figure}

Next we study the interplay between cutoff and twist conformation for the homogeneous chain with $d=\,0$. Keeping the same model parameters as above, the molecule unwinding is simulated  by increasing the helical repeat and  the appropriate $\bar{U}$ is determined for each $h$. The results are displayed in Fig.~\ref{fig:3}. The $P_j(R_0,\, t)$'s in 
Fig.~\ref{fig:3}(a) are computed by taking the $\bar{U}$'s obtained respectively from the plots in Fig.~\ref{fig:3}(b). As a main result, $\bar{U}$ markedly decreases for larger $h$. This is understood by observing that our helical chains are considered to be stable at room temperature also in the untwisted conformations although, in the latter, large amplitude base pair fluctuations would easily disrupt the hydrogen bonds and unstack the helix. Accordingly, helical molecules in a large $h$ conformation sustain only short scale fluctuations in order to preserve the overall stability. This result follows from the choice of a model which has been assumed to have no intrinsic stiffness i.e., $d=\,0$.  By further increasing $h$ the helix unwinds and tends to the ladder representation. Consistently, it is shown in Fig.~\ref{fig:3}(b), that the selected $\bar{U}$'s tend to the cutoff value determined for the PBD model \cite{io20}.

\begin{figure}
\includegraphics[height=7.0cm,width=8.0cm,angle=-90]{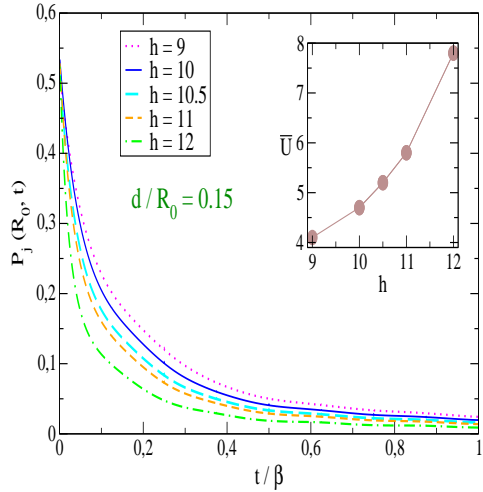}
\caption{\label{fig:4}(Color online)  
As in Fig.~\ref{fig:3}(a) but with a finite rise distance $d$.  Five twist conformations are considered. {The respective average twist angles $\bar \theta$  are (from top to bottom in the legend) \, $40^o, \,36^o, \, 34.3^o, \, 32.7^o, \, 30^o $.} For each $h$ value, the first passage probability is computed assuming the respective cutoff \={U} (in the inset) that fulfills the initial time condition.  }
\end{figure}

This trend is however reversed once we introduce a finite rise distance which provides a more realistic model for the helical molecule and accounts for the intrinsic stiffness of the chain. The results are shown in Fig.~\ref{fig:4} where the time dependent probability is computed as a function of time by varying the average twist angle. Now the finite $d$ confers stability to the helix which can sustain large amplitude base pair fluctuations also in the untwisted conformations. Accordingly the 
cutoff value, for which the zero time probability condition is fulfilled, grows versus $h$ as shown in the inset. For instance, given a chain with $h=\,10.5$ i.e. the standard DNA helical repeat at room temperature \cite{duguet93}, the obtained value $\bar{U} =\,5.2$ yields a maximum amplitude $\Lambda_{j}(T)=\,1.08 \AA$ for the first Fourier component in Eq.~(\ref{eq:04}). This, in turn, corresponds to a reasonable estimate of $\sim 2.2 \AA$ for the largest breathing fluctuation of the $j-th$ base pair with respect to the average helix diameter in the closed state. This length is $\sim 10 \%$ of $R_0$ and it is a fair measure for the threshold above which hydrogen bonds are disrupted \cite{pey08,manghi16}.

\subsection*{B. Sliding and bending}

\begin{figure}
\includegraphics[height=7.0cm,width=8.0cm,angle=-90]{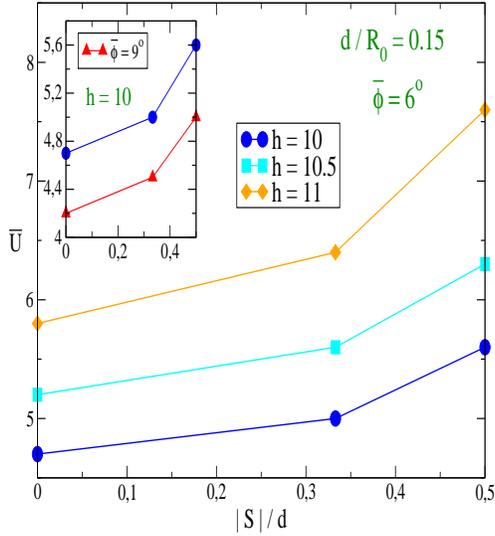}
\caption{\label{fig:5}(Color online) 
The cutoff \={U}, which enforces the constraint for the initial time probability, is plotted as a function of the ratio between absolute value of slide $S$ and bare rise distance $d$. The schematic of the molecule with sliding motion is given in Fig.~\ref{fig:1}(b).   Three twist conformations are considered while the average bending angle is 
$\bar{\phi}=6^{o}$.  The effect of a broader bending is shown in the inset for the conformation with $h=\,10$.  }
\end{figure}

Now we focus on the geometrical model of Fig.~\ref{fig:1}(b) which contains a further parameter, i.e. the sliding motion of neighboring base pairs having the overall effect to shorten the helix. The model has a finite rise distance and the sliding $S$ is tuned as a fraction of $d$ \cite{n3}. The probability in Eq.~(\ref{eq:03aa}) is computed also for this more complex configuration and the obtained cutoff $\bar{U}$ is plotted in Fig.~\ref{fig:5} as a function of the ratio $|S| / d$, assuming three twist conformations. As in the previous plots, the average bending angle is $\bar{\phi}=\,6^{o}$. 
It is found that $\bar{U}$ is enhanced in the presence of a finite $S$. This behavior holds for all helical conformations and in particular for the case \, $h=\,11$ with a sizeable sliding. The fact that a double helix with a pronounced base pair sliding should sustain broader radial fluctuations seems in accordance with the observation that  dsRNA is wider than dsDNA \cite{dekker14}. 

If however the short chain has a larger average bending flexibility, for instance due to melting bubbles formation or large angle deformations at specific dinucleotide steps \cite{widom04,archer06,vafa12}, then lower $\bar{U}$'s  should suffice in order to to fulfill the first-passage probability constraint. This is in fact the situation emphasized in the inset, whereby a chain with $\bar{\phi}=\,9^{o}$ is assumed for the twist conformation with $h=\,10$. Accordingly the calculated $\bar{U}$'s are somewhat reduced with respect to the case $\bar{\phi}=\,6^{o}$.

Altogether these findings are consistent with the interpretation that, at the microscopic level, the contraction of the rise distance $H_S$ due to the slide may confer an enhanced flexibility to the molecule which is then capable to sustain large scale fluctuations while maintaining the stable helical conformations. It is emphasized that our calculation takes $S$ and $h$ as independent variables whereas the structural parameters are intertwined in nucleic acids that is,  a larger slide is generally observed in helical molecules with a smaller average twist angle.
To meet these experimental observation, one may tune $S$ and set up a self-consistent procedure to compute, by free energy minimization, both radial cutoff and ensemble averaged helical repeat.

\begin{figure}
\includegraphics[height=7.0cm,width=8.0cm,angle=-90]{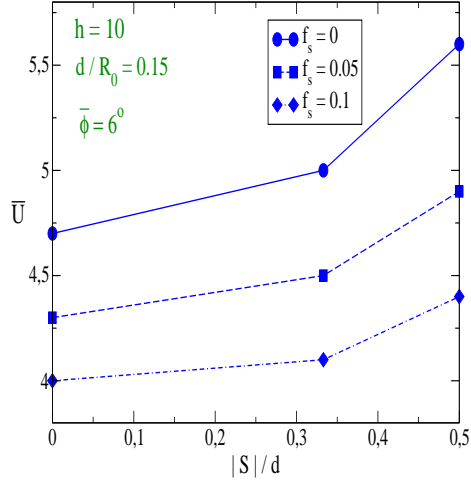}
\caption{\label{fig:6}(Color online) 
The cutoff \={U} is plotted as a function of the ratio between slide $S$ and bare rise distance $d$. The effect of the solvent potential in Eq.~(\ref{eq:01a}) is shown by tuning $f_s$ .  The twist conformation with $h=\,10$ is considered while the average bending angle is $\bar{\phi}=6^{o}$.    }
\end{figure}

Finally, we re-consider the Hamiltonian in Eq.~(\ref{eq:01}) and add the solvent term in Eq.~(\ref{eq:01a}) to the one-particle potential. By increasing $f_s$, one can simulate the effect of a higher salt concentration in the solvent which screens the electrostatic repulsion between complementary strands thereby enhancing the base pair dissociation energy hence, an intrinsically more stable molecule configuration \cite{albu14,singh15}. Assuming the twist conformation $h=\,10$ with
$\bar{\phi}=6^{o}$, we have re-run the program in order to test how the cutoff \=U varies in the presence of the solvent factor. The results are shown in Fig.~\ref{fig:6} where \=U is plotted as a function of $|S| / d$. In accordance to the expectation, \=U is reduced with respect to the model without solvent ($f_s=\,0$) and this holds both
in the case with and without sliding motion. For the intermediate sliding case \, $|S| / d =\,0.33$, we find a cutoff reduction of $\sim 18 \%$ for a energy barrier increase of $\sim 10 \%$ i.e., $f_s=\,0.1$.

\section*{V. Conclusions }

Analysis of the thermodynamics and flexibility of nucleic acids at the mesoscopic scale requires computation of a partition function obtained as an integral over the space configuration available for the base pair fluctuations. Focusing on a three dimensional Hamiltonian model for double stranded helical molecules, we have proposed a statistical method to select the integration cutoff on the amplitude of the base pairs separation, a crucial parameter for quantitative predictions of the physical properties. The method conceives the base pair thermal fluctuations as a Brownian motion for a particle subjected to the interactions which render the molecule stable hence, the base pair vibrates around the equilibrium helix diameter. Setting a \textit{zero time} condition for the particle trajectory, we have calculated the time dependent probability for the base pair to return to the initial position by summing over those particle histories which are physically consistent with the model potential. In turn, the first-passage probability obeys a \textit{zero time} condition which sets the constraint to establish the proper cutoff on the base pair fluctuations. Applying the method to a short homogeneous molecule, we find that the cutoff depends significantly on the macroscopic helical conformation defined by the three variables i.e. twist, bending and sliding for the relative motion of two base pairs in any dimer of the chain. In particular, for a realistic Hamiltonian model incorporating an intrinsic stiffness in the stacking potential, {the room temperature base pair fluctuations are estimated to be $\sim 10 \%$ of the helix diameter.} It is also found that the cutoff gets larger upon increasing the average twist angle in line with the expectation that the base pair fluctuations should be broader in untwisted helices.  Likewise, the radial cutoff grows in molecules with a significant slide which shortens the rise per base pair along the molecule axis. This seems consistent with the observation that A-form helices are indeed wider and shorter than B-helices in which the slide is small. These main conclusions hold in general and are not restricted to the chain here considered, although quantitative estimates of the cutoff may vary according to the set of parameters chosen to model hydrogen bonds, stacking forces and solvent environment for a helical molecule.  Thus, the first-passage probability method provides a sound statistical benchmark to select potential parameters and maximum base pair fluctuations for specific sequences and conformations of nucleic acids and, in particular, may foster mesoscopic studies of RNA properties so far less investigated than DNA.

\section*{ DATA AVAILABLITY STATEMENT}

The data that support the findings of this study are available from the corresponding author upon reasonable request.


\begin{thebibliography}{widest-label}

\bibitem{pablo07}
T.A. Knotts, N. Rathore, D.C. Schwartz,  J.J. de Pablo, \textit{J. Chem. Phys.} \textbf{126}, 084901 (2007).

\bibitem{doye11}
T.E. Ouldridge, A.A. Louis, J.P.K. Doye,  \textit{J. Chem. Phys.} \textbf{134}, 085101 (2011).

\bibitem{jeon14}
J.-H. Jeon, W. Sung, \textit{J. Biol. Phys.}  \textbf{40}, 1-14 (2014).

\bibitem{bish14}
C. Nisoli, A.R. Bishop, \textit{J. Chem. Phys.} \textbf{141}, 115101 (2014).

\bibitem{liang14}
H. Li, Z. Wang, N. Li, X. He, H. Liang, \emph{J. Chem. Phys.} \textbf{141},  044911 (2014).

\bibitem{pey93}
T. Dauxois, M. Peyrard, A.R. Bishop,  \emph{Phys. Rev. E}  \textbf{47},   R44-R47  (1993).


\bibitem{campa98}
A. Campa, A. Giansanti, \textit{Phys. Rev. E}  \textbf{58}, 3585-3588  (1998).

\bibitem{zdrav06}
S. Zdravkovi\'{c},  M.V. Satari\'{c},  \emph{Phys. Rev. E}  {\bf 73},   021905  (2006).

\bibitem{bish09}
B.S. Alexandrov, V. Gelev, Y. Monisova, L.B. Alexandrov, A.R. Bishop, K.$\varnothing$. Rasmussen, A. Usheva,  \textit{Nucleic Acids Res.} {\bf 37}, 2405-2410 (2009).

\bibitem{singh09}
S. Srivastava, Y. Singh  \textit{EPL} \textbf{85}, 38001 (2009).

\bibitem{io09}
M. Zoli,   \emph{Phys.Rev. E}   \textbf{79},  041927 (2009).

\bibitem{io10}
M. Zoli,   \emph{Phys.Rev. E}   \textbf{81},  051910 (2010).

\bibitem{singh11}
S. Srivastava, N. Singh,  \emph{J. Chem. Phys.} \textbf{134},   115102  (2011).

\bibitem{falo12}
R. Tapia-Rojo, D. Prada-Gracia, J.J. Mazo, F. Falo, \emph{Phys.Rev. E}    \textbf{86},  021908  (2012).

\bibitem{hando12}
A. Sulaiman, F.P. Zen, H. Alatas,  L.T. Handoko, \textit{Phys. Scr.} \textbf{86},  015802 (2012).

\bibitem{albu14}
D.X. Macedo, I. Guedes, E.L. Albuquerque,    \textit{Physica A}  \textbf{404}, 234-241 (2014).

\bibitem{io14a}
M. Zoli,  \textit{Soft Matter} {\bf 10},  4304-4311 (2014).

\bibitem{lak19}
I.V. Likhachev, V.D. Lakhno, \textit{Chem. Phys. Lett.} \textbf{727},  55-58 (2019).

\bibitem{kalos20}
M. Hillebrand, G. Kalosakas,  Ch. Skokos,  A. R. Bishop, \emph{Phys. Rev. E} \textbf{102}, 062114 (2020).

\bibitem{weber13}
G. Weber,   \emph{Nucleic Acids Res.}  \textbf{41}, e30 (2013).

\bibitem{weber19}
E. de Oliveira Martins, V.B. Barbosa, G. Weber,   \emph{Chem. Phys. Lett.}  \textbf{715}, 14-19 (2019).

\bibitem{pey08}
M. Peyrard, S. Cuesta-L\'{o}pez, G. James, \textit{Nonlinearity} \textbf{21}, T91 (2008).

\bibitem{gueron85}
M. Gu\'{e}ron, J.-L. Leroy,  in \textit{Methods in Enzymology} \textbf{261}, Academic Press, San Diego
(1985).

\bibitem{pey09}
M. Peyrard, S. Cuesta-L\'{o}pez, D. Angelov, \textit{J. Phys.: Condens. Matter} {\bf 21}, 034103  (2009).


\bibitem{erp18}
M. Marty-Roda, O. Dahlen, T.S. van Erp, S. Cuesta-L\'{o}pez, \textit{Phys. Biol.} \textbf{15}, 066001 (2018).

\bibitem{coll95}
F. Zhang, M.A. Collins, \textit{Phys. Rev. E} {\bf 52}, 4217 (1995).

\bibitem{croq00}
T. Strick, J.-F. Allemand, V. Croquette, D. Bensimon, \textit{Prog. Biophys. Mol. Biol.} \textbf{74},  115–140 (2000).

\bibitem{maiti15}
A. Garai,  S. Saurabh,  Y. Lansac, P.K. Maiti, \textit{J. Phys. Chem. B}  \textbf{119}, 11146-11156 (2015).


\bibitem{joy05}
M. Joyeux, S. Buyukdagli,  \textit{Phys.\ Rev.\ E} {\bf 72}, 051902 (2005).


\bibitem{ares05}
S. Ares, N.K. Voulgarakis, K.{\O}. Rasmussen, A.R. Bishop, \textit{ Phys.\ Rev.\ Lett.}   {\bf 94}, 035504 (2005).

\bibitem{zhang97}
Y.L. Zhang, W.M. Zheng, J.X. Liu,  Y.Z. Chen,   \textit{ Phys. Rev. E}  \textbf{56},  7100-7115  (1997).

\bibitem{pey95}
T. Dauxois, M. Peyrard,  \emph{Phys. Rev. E}  \textbf{51},   4027  (1995).

\bibitem{wart85}
R.M. Wartell, A.S. Benight, \textit{Phys. Rep.} \textbf{126},   67-107 (1985).


\bibitem{singh15}
A. Singh, N. Singh,  \emph{Physica A} \textbf{419}, 328-334  (2015).


\bibitem{io11a}
M. Zoli,  \emph{Eur. Phys. J. E} {\bf 34}, 68 (2011).

\bibitem{io14c}
M. Zoli,  \textit{J. Theor. Biol.} {\bf 354},  95-104 (2014).

\bibitem{klein04}
H. Kleinert,  {\it Path Integrals in Quantum Mechanics, Statistics,
Polymer Physycs and Financial Markets}, World Scientific Publishing, Singapore (2004).

\bibitem{io16b}
M. Zoli,   \textit{J. Chem. Phys.}   {\bf 144},   214104  (2016). 

\bibitem{io18c}
M. Zoli,       \emph{J. Chem. Phys.}  \textbf{148},   214902 (2018).

\bibitem{io18}
M. Zoli,   \textit{Physica A} \textbf{492}, 903-915 (2018).

\bibitem{io18a}
M. Zoli, \textit{EPL} {\bf 123},   68003 (2018).

\bibitem{io18b}
M. Zoli, \textit{EPL} {\bf 130}, 28002 (2020).

\bibitem{io16a}
M. Zoli, \textit{Phys. Chem. Chem. Phys.}   {\bf 18}, 17666  (2016).


\bibitem{io20}
M. Zoli,  \textit{Phys. Chem. Chem. Phys.} {\bf 22},  26901 (2020).

\bibitem{io11b}
M. Zoli,       \emph{J. Chem. Phys.}  \textbf{135},    115101 (2011).

\bibitem{calla}
C.R. Calladine, H.R. Drew,   \emph{Understanding DNA}, Academic Press, San Diego, USA, 1992.

\bibitem{gonz13}
E. Herrero-Gal\`{a}n, M.E. Fuentes-Perez,  C. Carrasco,  J.M. Valpuesta, J.L. Carrascosa,  F. Moreno-Herrero,  J.R. Arias-Gonzalez, \textit{J. Am. Chem. Soc. } \textbf{135}, 122-131 (2013).

\bibitem{dekker14}
J. Lipfert, G.M. Skinner, J.M. Keegstra, T. Hensgens, T. Jager, D. Dulin,
M. K\"{o}ber, Z. Yu, S.P. Donkers, F.-C. Chou, R. Das, and N.H. Dekker, \textit{Proc. Natl. Acad. Sci. U.S.A.} {\bf 111}, 15408-15413 (2014).

\bibitem{olson03}
X.-J. Lu, W.K. Olson, \emph{Nucleic Acids Res.}   \textbf{31}, 5108-5121  (2003). 

\bibitem{wang79}
J.C. Wang,   \emph{Proc. Natl. Acad. Sci. USA}   \textbf{76},  200-203 (1979).

\bibitem{n3}
The slide is conventionally defined negative if the upper base pair in the dimer shifts to the right with respect to the lower base pair \cite{calla}. 


\bibitem{oroz10}
I. Faustino, A. P\'{e}rez,  M. Orozco, \textit{ Biophys. J.} \textbf{99}, 1876-1885 (2010).

\bibitem{lavery14}
M. Pasi, J.H. Maddocks, D. Beveridge, T.C. Bishop, D.A. Case, T. Cheatham, P.D. Dans, B. Jayaram, F. Lankas, C. Laughton,
J. Mitchell, R. Osman, M. Orozco, A. P\'{e}rez, D. Petkevi\v{c}i\={u}t\.{e}, N. Spackova, J. Sponer, K. Zakrzewska and
R. Lavery, \emph{Nucleic Acids Res.}   \textbf{42}, 12272-12283  (2014). 


\bibitem{dick89}
R.E. Dickerson,   \textit{Nucleic Acids Res.}  \textbf{17}, 1797-1803 (1989).

\bibitem{olson95}
A.A. Gorin,  V.B. Zhurkin, W.K. Olson,  \textit{J. Mol. Biol.} \textbf{247}, 34-48 (1995).




\bibitem{kovac73}
M. Fixman, J.J. Kovac, \textit{J. Chem. Phys.} \textbf{58}, 1564 (1973).

\bibitem{busta92}
S. Smith, L. Finzi, C. Bustamante,  \emph{Science}   \textbf{258},  1122-1126 (1992).

\bibitem{barbi99}
M. Barbi, S. Cocco, M. Peyrard, \textit{Phys. Lett. A }\textbf{253},  358 (1999).


\bibitem{weber06}
G. Weber, \textit{Europhys. Lett.} \textbf{73}, 806 (2006).


\bibitem{proho93}
Y.Z. Chen, E.W. Prohofsky, \textit{Phys.\ Rev.\ E} {\bf 47}, 2100 (1993).

\bibitem{olson10}
G. Zheng, L. Czapla, A.R. Srinivasan,  W.K. Olson, \textit{Phys. Chem. Chem. Phys.}  \textbf{12}, 1399-1406 (2010).

\bibitem{metz11}
S. Talukder, P. Chaudhury, R. Metzler, and S.K. Banik, \emph{J. Chem. Phys.} \textbf{135},  165103 (2011).

\bibitem{maher13}
J.P. Peters, S.P. Yelgaonkar, S.G. Srivatsan, Y. Tor, L.J. Maher III,   \emph{Nucleic Acids Res.}   \textbf{41}, 10593-10604  (2013). 

\bibitem{onuf19}
A.V. Drozdetski, A. Mukhopadhyay,  A.V. Onufriev, \textit{Front. Phys.} \textbf{7}, 195 (2019).

\bibitem{druk01}
K. Drukker, G. Wu, G.C. Schatz,   \emph{J. Chem. Phys.} \textbf{114},  579-590  (2001).


\bibitem{io14b}
M. Zoli,  \textit{J. Chem. Phys.}  {\bf 141}, 174112 (2014).


\bibitem{io12}
M. Zoli, \textit{J. Phys.: Condens. Matter} {\bf 24},  195103  (2012).


\bibitem{joy07}
S. Buyukdagli, M. Joyeux, \textit{Phys.\ Rev.\ E} {\bf 76}, 021917 (2007).

\bibitem{zgarb14}
M. Zgarbov\'{a}, M. Otyepka, J. Sponer, F. Lankas,  P. Jurecka,  \textit{J. Chem. Theory Comput.} \textbf{10}, 3177-3189 (2014).

\bibitem{io17}
M. Zoli, \textit{J. Phys.: Condens. Matter} {\bf 29},     225101 (2017).

\bibitem{io19}
M. Zoli, \textit{Phys. Chem. Chem. Phys.}   {\bf 21},   12566  (2019).

\bibitem{bonnet03}
G. Altan-Bonnet, A. Libchaber, O. Krichevsky,   \emph{Phys. Rev. Lett.} \textbf{90},   138101 (2003).

\bibitem{marcus13}
C. Phelps, W. Lee, D. Jose, P.H. von Hippel,  A.H. Marcus, \textit{Proc. Natl. Acad. Sci. U.S.A.}   \textbf{110}, 17320-17325 (2013).

\bibitem{stasiak97}
K. Kiianitsa, A. Stasiak, \textit{Proc. Natl. Acad. Sci. U.S.A.}   \textbf{94}, 7837-7840, (2013).

\bibitem{kalos11}
A. Apostolaki, G. Kalosakas, \textit{Phys. Biol.}  \textbf{8}, 026006 (2011).  

\bibitem{wang12}
J.L. Killian, M. Li, M.Y. Sheinin, M.D. Wang,   \textit{Curr. Opin. Struct. Biol.}  \textbf{22}, 80–87 (2012).

\bibitem{marko15}
J.F. Marko,  \textit{Physica A}  \textbf{418}, 126-153 (2015).

\bibitem{lee19}
S.-R. Choi, N.-H. Kim, H.-S. Jin, Y.-J. Seo, J. Lee, J.-H. Lee, \textit{Comput. Struct. Biotech. J.} \textbf{17},  797-804 (2019).

\bibitem{fehi}
R.P. Feynman,  A.R. Hibbs,   {\it Quantum Mechanics and Path Integrals}, (Mc Graw-Hill, New York,  1965).


\bibitem{nn1}
In Transfer Integral methods applied to the PBD model, the partition function is customarily obtained by applying periodic boundary conditions which amount to close the linear chain into a loop \cite{joy05,kalos20}. This procedure may be questionable in short chains due to the relevance of finite size effects. This drawback is solved by the Path Integral method which imposes the closure condition on the time axis whereas the chain maintains the open ends in real space.  {Furthermore, the PI method can be also extended to deal with circular molecules \cite{io13}. }

\bibitem{io13}
M. Zoli, \textit{J. Chem. Phys.}  {\bf 138}, 205103 (2013).

\bibitem{maj05}
S.N. Majumdar, \textit{Curr. Sci.} \textbf{89}, 2076 (2005).

\bibitem{n2}
While the Fourier coefficients are integrated on an even domain, too negative $a_m$'s are discarded due to the physical condition associated to the hard core of the one particle potential, as discussed after Eq.~(\ref{eq:01}). Such condition depends on the parameter which regulates the range of the Morse potential. This asymmetry in the choice of $a_m$'s included in the computation explains why $P_j(R_0,\, 0)$ may get slightly larger than $1/2$. Hence the approximation sign used in the text.
 

\bibitem{kim14}
T.T. Le, H.D. Kim,    \emph{Nucleic Acids Res.}   \textbf{42},  10786-10794  ( 2014).


\bibitem{eijck11}
L. van Eijck, F. Merzel, S. Rols, J. Ollivier, V.T. Forsyth, M.R. Johnson,  \textit{Phys. Rev. Lett.}  \textbf{ 107},  088102 (2011).

\bibitem{weber15}
I. Ferreira, T.D. Amarante,  G. Weber,    \emph{J. Chem. Phys.} \textbf{143}, 175101 (2015).

\bibitem{berg06}
D. Andreatta, S. Sen, J.L. P\'{e}rez Lustres, S.A. Kovalenko, N.P. Ernsting, C.J. Murphy, R.S. Coleman, M.A. Berg, \textit{J. Am. Chem. Soc.}   \textbf{128}, 6885-6892 (2006).

\bibitem{duguet93}
M. Duguet,   \emph{Nucleic Acids Res.}  \textbf{21}, 463-468 (1993).

\bibitem{manghi16}
M. Manghi, N. Destainville, \textit{Phys. Rep.} \textbf{631}, 1 (2016).


\bibitem{widom04}
T.E. Cloutier, J. Widom,  \emph{Mol. Cell}  \textbf{ 14}, 355-362 (2004).

\bibitem{archer06}
C. Yuan, E. Rhoades, X.W. Lou, L.A. Archer, \emph{Nucleic Acids Res.}   \textbf{34},  4554-4560  (2006).

\bibitem{vafa12}
R. Vafabakhsh, T. Ha,  \emph{Science}  \textbf{337},   1097-1101   (2012).






 
 

 











\end{thebibliography}
\end{document}